\documentclass{article}
\usepackage{times}
\usepackage{hyperref}
\usepackage{url}
\usepackage{graphicx}
\usepackage{float}
\usepackage[caption = false]{subfig}
\usepackage{url}
\usepackage{multirow}

\usepackage[margin=1.3in]{geometry}

\usepackage{amsmath}

\newtheorem{myTheo}{Theorem}

\usepackage{amssymb}
\usepackage{algorithm}
\usepackage{algorithmic}

\newcommand\Co{\text{Conf}}

\title{Confidence Statements for Ordering Quantiles}

\author{Carlos A. de B. Pereira\\
Institute of Mathematics and Statistics\\
University of S\~ao Paulo, S\~ao Paulo, Brazil\\
{\it cpereira@ime.usp.br}
\and
Cassio P. de Campos\\
Dalle Molle Institute for Artificial Intelligence\\
Manno, Switzerland\\
{\it cassio@idsia.ch}
\and
Adriano Polpo\\
Federal University of S\~ao Carlos\\
S\~ao Carlos, Brazil\\
{\it polpo@ufscar.br}}

\begin{document}
\date{June, 2014} 
\maketitle

\begin{abstract}
  This work proposes {\it Quor}, a simple yet effective nonparametric
  method to compare independent samples with respect to corresponding
  quantiles of their populations. The method is solely based on the
  order statistics of the samples, and independence is its only
  requirement.  All computations are performed using
  exact distributions with no need for any asymptotic considerations,
  and yet can be run using a fast quadratic-time dynamic programming
  idea. Computational performance is essential in high-dimensional domains, such
  as gene expression data. We describe the approach and discuss on the
  most important assumptions, building a parallel with assumptions and
  properties of widely used techniques for the same problem.
  Experiments using real data from biomedical studies
  are performed to empirically compare Quor and other methods in a
  classification task over a selection of high-dimensional data sets.
\end{abstract}

\section{Introduction}
\label{sec:intro}

A common procedure in Statistics and Machine Learning when dealing
with data sets of thousands of variables is to {\it sort} all these
variables according to some measure that identifies how important they
are to predict and/or retrospectively understand a certain target
variable (or equivalently an indicator that tells in which group or
{\it population} belongs each sample). Classical examples of such a
procedure are the Student's $t$-test and the Wilcoxon's rank-sum
$u$-test~\cite{dems,fay,utest}, whose statistics are often used to
sort variables into some order of importance.  Arguably, they
represent the most commonly used methods for this problem in
biomedical applications, in part because of their prompt availability
and easiness of use. A typical scenario is to have gene expression
data of cancer patients, and a {\it class} variable that identifies
whether the patient {\it relapsed} or not (in other word, whether the
cancer came back after treatment/surgery or not). The ability to sort
variables in some meaningful order has a range of applications in many
fields, and can also be seen as means of performing {\it feature selection}~\cite{mitchell,witten}.

This work describes a simple yet effective approach, named {\it Quor},
to sort variables according to the order relationship of arbitrary
quantiles of the variable's distributions under different groups. The
method computes a value that indicates the confidence that such
quantiles of these distributions are sorted in some pre-defined
way. As example, suppose we have two populations and are interested in
the median values of a variable representing the level of expression
of a particular gene. The goal is to obtain the confidence that ``the
median expression of that gene in the first population is strictly
smaller (or greater) than the median expression of the same gene in the
second population''.  The comparison of medians might suggest that the
gene is under-expressed, over-expressed, or simply that there is no
significant difference of expression between the populations. As other
methods with similar purpose, Quor can be used as a first aid for the
later application of other sophisticated statistical or biological
procedures: Its simplicity may avoids expensive and time-consuming
analyses of uninteresting variables, or at least may help to
prioritize further analyses in order to save valuable time and
biological materials/machinery.

Methods for this problem may rely on eventually unrealistic
hypotheses, such as normality of samples, asymptotic behavior of some
statistics, reasonably large-sized samples, approximate computations,
comparisons of only two populations, need of equivalent number of
samples in the groups (across distinct variables), necessity of
multiple-test corrections (to avoid encountering significant results
{\it by chance}), no evidence for the null hypothesis, among
others~\cite{raftery}.  These issues might be aggravated by having
data sets with only a few patients and a large number of genes. On the
other hand, Quor is nonparametric and assumes nothing but independence
of samples. It can deal with different number of samples and missing
data, and yet can properly compare these variables; all computations
are performed exactly, without any asymptotic assumption. Moreover,
its computations can be carried on in quadratic time in the number of
samples, which is (roughly) as fast as other methods based on
$t$-tests and $u$-tests. Other approaches for median test do exist (we
refer to Chapter 6 of~\cite{Gibbons2003} for some examples), but they
either fail in at least one of the issues mentioned previously in this
paragraph, or are not able to order arbitrary quantiles of (arbitrary)
many populations.

This paper is divided as follows. Section~\ref{sec2} describes the
approach in details and presents its computational
complexity. Sections~\ref{sec4} and~\ref{sec5} analyze real
biomedical data sets and compare the empirical performance of the
methods. Section~\ref{sec6} concludes the paper and discuss on future
work.

\section{Unveiling Quor}
\label{sec2}

We describe here the details of Quor and present an efficient
algorithm for its computation. The method is built on the ideas of
confidence statements developed long ago~\cite{basu1981,kiefer1977}
and revisited more recently~\cite{zellner2004}. The proposed method
uses nonparametric confidence intervals for quantiles based on the
binomial distribution \cite{degroot1975}.  Its goal is to compute a
confidence value indicating how much one believes that quantile
parameters of different populations/groups are ordered among themselves. We do not
assume any particular quantile nor a specific number of populations,
even though the case of comparing medians of two populations is
arguably the most common scenario for its application.

The problem is defined as follows. Let $Q_1,\ldots,Q_n$ represent the
quantiles at arbitrary percentages $q_1,\ldots,q_n$, respectively, for
$n$ populations, and let ${\bf x}_i=(x_i^{(1)},\ldots,x_i^{(m_i)})$,
for $i=1,\ldots,n$, be data samples from those populations, where the
sample from population $i$ has
size $m_i$. The goal is to produce a confidence value in $[0,1]$ that
indicates how much we believe in the statement $Q_{i_1}<Q_{i_2}<\ldots
<Q_{i_n}$, where $(i_1,\ldots,i_n)$ is a permutation of
$(1,\ldots,n)$. As in other nonparametric methods \cite{wasserman2006},
the order and the values of the tuples of order statistics will be the
solely observations needed.

Consider an event $A$ related to a random variable $X_i$ whose quantile at $q_i$
is $Q_{i}$. $\Pr(A\mid Q_{i})$ indicates the probability of the event $A$ with $Q_{i}$ 
known. The quantile $Q_{i}$ of $X_i$ is a population parameter that
satisfies the following inequality: 
\begin{equation}
\Pr(\{X_i \leq Q_{i}\} \mid Q_{i})\geq q_i,
\end{equation}
\noindent while in the continuous case, this inequality is tight.

Let $(X_i^{(1)},X_i^{(2)},\ldots,X_i^{(m_i)})$ be the ordered vector of $m_i$
independent and identically distributed continuous random
variables. Since the probability of one observation being smaller than
the population quantile $Q_{i}$ is $q_i$, it is straightforward that for
the $j$th $(j=1,2,\ldots,m_i)$ order statistics, $X_i^{(j)}$, the
following probability expression holds:
\begin{equation}
\Pr(\{X_i^{(j)} \leq Q_{i}\} \mid Q_{i}) = \sum\limits_{k=j}^{m_i}
\binom{m_i}{k} q_i^k(1-q_i)^{m_i-k},
\label{e1}
\end{equation}
\noindent and symmetrically
\begin{equation}
\Pr(\{X_i^{(j)} \geq Q_{i}\} \mid Q_{i}) = \sum\limits_{k=0}^{j-1}
\binom{m_i}{k} q_i^k(1-q_i)^{m_i-k}.
\label{e2}
\end{equation}
\noindent
These equalities come from probabilities obtained with a binomial distribution with sample size $m_i$ and
probability of success $q_i$.  Consider a sequence of pairs of
order statistics ($(X_1^{(j_1)}$, $X_1^{(j'_1)})$,
$(X_2^{(j_2)};X_2^{(j'_2)})$, $\ldots$, $(X_n^{(j_n)};X_n^{(j'_n)})$), each
of them chosen from the $i$th group (with abuse of notation, let first and last groups
have $X_1^{(j_1)}=-\infty$ and $X_n^{(j'_n)}=\infty$, respectively),
and consider the event $\text{E}$ as follows.
\[
\text{E} = \bigcap_{i=1}^n \{X_i^{(j_i)} \leq Q_{i}\leq  X_i^{(j'_i)} \}
=\bigcap_{i=1}^n \left(\{X_i^{(j'_i)} \geq Q_{i}\} \setminus \{ X_i^{(j_i)} >
Q_{i}\}\right).
\]
\noindent
Given the independence among the samples, one can
compute $\Pr(\text{E}\mid Q_{1},\ldots,Q_{n})=$
\begin{equation}
\prod_{i=1}^n \max\left\{0;\left[
\Pr(\{X_i^{(j'_i)} \geq Q_{i}\} \mid Q_{i})-\left(1-
\Pr(\{X_i^{(j_i)} \leq Q_{i}\} \mid Q_{i})\right) \right]\right\},
\label{eq:confidence}
\end{equation}
using the product of binomial probabilities from Equations~\eqref{e1} and~\eqref{e2}.

Let ${\bf x}_i$ be sorted, for every $i$. After these samples are
observed, the only unknown quantities of interest are the quantiles
$Q_{i}$ of the populations being studied. By replacing random
variables with their observations, we create the statement $\text{e}:$
\begin{equation}
\text{e}=\bigcap_{i=1}^n \{x_i^{(j_i)} \leq Q_{i}\leq x_i^{(j'_i)}\},
\label{eq:e}
\end{equation}
\noindent which has confidence given by
Expression~\eqref{eq:confidence}. Note that we only need the orders
$j_i$ and $j'_i$ for every $i$, and not the actual observed values of
$X_i$, for computing with Expression~\eqref{eq:confidence}, that is,
the confidence of $\text{e}$ could be equivalently defined as the
confidence of $((0;j'_1);(j_2;j'_2);\ldots;(j_n;1+m_n))$, where we
conveniently define that, for every $i=1,\ldots,n$,
$x_i^{(j)}=-\infty$ for every $j<1$ and $x_i^{(j)}=\infty$ for every
$j>m_i$. Hence, we might say that $\text{e}$ is represented by the
list $((0;j'_1);(j_2;j'_2);\ldots;(j_n;1+m_n))$.  At this point after
sampling, we call the value of Expression~\eqref{eq:confidence} a
confidence value instead of probability value, in order to keep
terminology precise. This confidence regards the unknown quantities of
interest, in our case the parameters $Q_{i}$.  The idea is to look for
statements $\text{e}$ that are able to tell us something about
$Q_{i}$. Without loss of generality, assume that we take a list of
pairs of order statistics $((0;j'_1);(j_2;j'_2);\ldots;(j_n;1+m_n))$
such that, in the observed sets, we have $x_i^{(j_i)}
<x_i^{(j'_i)}<x_{i+1}^{(j_{i+1})}$, for every $1\leq i <n$ (we do not
lose generality because in case we want to sort these observations in
some other order, we simply rename the variables). With this fact and
a quick analysis of Expression~\eqref{eq:e}, we derive
\begin{equation}
\left(\forall_{1\leq i <n}:~ x_i^{(j_i)} <x_i^{(j'_i)}<x_{i+1}^{(j_{i+1})}\right)
\land \text{e} \Rightarrow \bigcap_{i=1}^{n-1} \{Q_{i} <
Q_{{i+1}}\},
\label{eq:xx}
\end{equation}
\noindent that is, the assertion in the left-hand side of
Expression~\eqref{eq:xx} implies an order for the quantiles, so its
confidence is a lower bound for the confidence of the right-hand
side. Because we know how to compute the confidence value of
$\text{e}$ through Expression~\eqref{eq:confidence}, and any time 
the assertion $x_i^{(j_i)}<x_i^{(j'_i)}<x_{i+1}^{(j_{i+1})}$ is false
for any $i$ the result of Expression~\eqref{eq:confidence} becomes zero,
we have the following relation.
\begin{equation}
\Co(\text{e}) \leq \Co\left(\bigcap_{i=1}^{n-1} \{Q_{i} <
Q_{{i+1}}\}\right).
\label{eq:ac}
\end{equation}
In order to compute the best possible lower bound for the confidence
value in the right-hand side of Expression~\eqref{eq:ac}, we run over all tuples
$((0;j'_1);(j_2;j'_2);\ldots;(j_n;1+m_n))$ of orders that comply
with the linear order
\[
x_{1}^{(0)} <x_{1}^{(j'_{1})}
<x_{2}^{(j_{2})}<x_{2}^{(j'_{2})}< \ldots <
x_{n}^{(j_{n})}< x_{n}^{(1+m_n)}
\]
\noindent and keep the maximum confidence value of the $\text{e}$
statements built from these tuples as our estimation for the
confidence that $Q_{1}<Q_{2}<\ldots <Q_{n}$ holds true. Because we
want to maximize such confidence, we will always choose $j_i$ such
that $x_i^{(j_i)}$ is the smallest possible value greater than
$x_{i-1}^{(j'_{i-1})}$, that is, the value of $j_i$, in order to
maximize the confidence, is uniquely computable from the value of
$j'_{i-1}$ (that is because there is no reason to leave a larger gap
between them if a smaller gap is possible, as smaller gaps will yield
higher confidence values). Hence, we can ease on the terminology and
compute the confidence of $\text{e}$ by $\Co(j'_1,\ldots,j'_{n-1})$,
as these values are enough to find all $j_2,\ldots,j_n$ (that lead to
the highest possible confidence) and then to use
Expression~\eqref{eq:confidence} to obtain the confidence value. For
easy of computation, we define the confidence using a sum of
logarithms.
\begin{eqnarray}
\Co_i(j'_{i-1},j'_{i}) = \log\left(\max\left\{0;\Pr(\{X_i^{(j'_i)}
  \geq Q_{i}\} \mid Q_{i})+\Pr(\{X_i^{(j_i)} \leq Q_{i}\} \mid Q_{i})
  - 1\right\}\right),
\label{eq:cok}
\end{eqnarray}
\noindent where $j_i$ is obtained from $j'_{i-1}$ (by looking the
data) as just explained, and $\log 0$ is $-\infty$. Note that the value of
$\Co_i(j'_{i-1},j'_{i})$ will equal to $-\infty$ whenever the values
$j'_{i-1},j'_{i}$ do not induce a ``valid'' order in the observations,
that is, whenever there is no element $x_i^{(j_i)}$ strictly inside
$[x_{i-1}^{(j'_{i-1})},x_i^{(j'_i)}]$.  We are interested in
$\exp(\sum_{i=1}^n\Co_{i}(j'_{i-1},j'_{i}))$, which is our statistical
conclusion based on the observed samples that the quantiles (of the
populations) are ordered according to the permutation
$(1,\ldots,n)$. This procedure is presented in Algorithm~\ref{algo1},
which is explained in the continuation. We recall that if one wants to
compute the confidence of some other order, for instance
$Q_{i_1}<Q_{i_2}<\ldots <Q_{i_n}$, where $(i_1,\ldots,i_n)$ is a
permutation of $(1,\ldots,n)$, then they simply need to rename the
variables accordingly before invoking the algorithm, while if one
wants to check for every possible order of the quantiles, there would
be $n!$ permutations to check, which is fast for small values of $n$
(the vast majority of biomedical studies has $n\leq 6$ groups).
\begin{algorithm}
\caption{{\bf (Quor Core)} Finding the confidence value of a statement about the
  ordering of a quantile parameter among $n$ populations.}
\label{algo1}
\begin{algorithmic}
\item[\textbf{Input}]a data set with samples ${\bf
    x}_i=(x_i^{(1)},\ldots,x_i^{(m_i)})$, for $i=1,\ldots,n$, and the
  quantiles of interest $q_i$, for each $i$.
\item[\textbf{Output}]the log-confidence value that the
  statement $Q_{1}<Q_{2}<\ldots <Q_{n}$ holds.
\item[1] For every $i$ in $1,\ldots,n$, sort ${\bf x}_i$.
\item[2] Pre-compute the values that appear in
  Equations~\eqref{e1} and \eqref{e2} by making a cache to be used in
  the computation of Equation~\eqref{eq:cok}, for every
  $i=1,\ldots,n$: $\text{cache}(i,0)\leftarrow (1-q_i)^{m_i}$ and for $j=1,\ldots,m_i$:
\[
\text{cache}(i,j) \leftarrow \text{cache}(i,j-1) + \binom{m_i}{j} q_i^j(1-q_i)^{m_i-j}.
\]
\item[3] Let $D_i$ be a vector of size $m_i$ (defined from $1$ to $m_i$),
  for each $i=1,\ldots,n-1$. Initialize
  $D_1[\ell'_1]\leftarrow \Co_1(0,\ell'_{1})$, for every
  $1\leq\ell'_1\leq m_1$. If $n=2$, then go to line 5.
\item[4] For $i=2,\ldots,n-1$, do:
\item[\quad ~ ~] For $\ell'_i=1,\ldots,m_i$, do:\\
\[ \quad \quad ~ D_i[\ell'_i] = \max_{1\leq \ell'_{i-1}\leq m_{i-1}} (D_{i-1}[\ell'_{i-1}] +
\Co_i(\ell'_{i-1},\ell'_{i})). \]\\
\item[5] Return $\max_{1\leq \ell'_{n-1}\leq m_{n-1}} (D_{n-1}[\ell'_{n-1}] +\Co_n(\ell'_{n-1},1+m_n))$.
\end{algorithmic}
\end{algorithm}
\begin{myTheo}
Algorithm~\ref{algo1} uses space $O(n + m)$ and takes time $O(m\log m)$ if
$n=2$, and $O(m\max_i m_i)$ otherwise. 
\end{myTheo}
\noindent {\it Proof.} Step 1 needs to sort each of the samples ${\bf
  x}_i$, which takes $O(m_i\log m_i)$ for each $i$, so in total
$O(m\log\max_i m_i)$. Step 2
pre-computes the partial binomial sums. By doing it in a proper order,
this can be accomplished in constant time for each
$\text{cache}(i,j)$, and the loop will execute $O(\sum_i m_i)=O(m)$
times.  Step 3 initializes the data structures $D_i$ for the dynamic
programming. Summed altogether, they spend $O(m)$ space and $O(m)$
time to initialize. Step 4 performs the dynamic programming. The
number of times both loops are executed altogether is $O(m)$. The
internal maximization takes less than $O(\max_i m_i)$, so this step has time
$O(m\max_i m_i)$ in the worst case (and it is not run for $n=2$). Finally, Step 5 takes time
$O(m)$. $\blacksquare$

Algorithm~\ref{algo1} is very fast. First, many common practical cases
have $n=2$. In this case, the Step 4 of the algorithm is skipped and
the whole algorithm runs in linear time (except for the sorting of Step
1).  Second, this worst-case time complexity is pessimistic, and the
algorithm will usually run in sub-quadratic time. Third,
$m$ tends to be a reasonable small number (for example, biomedical
data sets hardly contain more than one hundred patients).

The correctness of Algorithm~\ref{algo1} comes from its simple dynamic
programming formulation. At each main loop $i$ of Step 4, we have
already solved the problem up to $i-1$ groups in the best possible way,
for each one of the positions one could choose for the last
placeholder $\ell'_{i-1}$, which we saved in
$D_{i-1}[\ell'_{i-1}]$. The choices of the yet-to-decide values
$\ell'_i,\ldots,\ell'_n$ do not depend on the positions of the
placeholders before $\ell'_{i-1}$, only on $\ell'_{i-1}$ itself. Thus, the dynamic
programming does the job of increasingly building new solutions. This
step is inspired by other dynamic programming algorithms, such as
those for $k$-means in one dimensional vectors~\cite{kmeans}, yet
different because of the nature of our confidence function to be
optimized. It is worth noting that some computations in
Algorithm~\ref{algo1} may suffer from numerical problems. We have
implemented it using incomplete Beta functions and arbitrary precision
when needed; this is usually the case if $m$ is reasonably large
(greater than some hundreds).

The confidence value obtained with Quor can be used in procedures that
want to optimize their utility functions, because they can provide
information about differences in quantiles as well as similarities in
quantiles (in the case that confidences are low in both
directions). This is in contrast to usual hypothesis testing, where no
evidence in favor of the null hypothesis can be obtained. A possible
negative characteristic of Quor is that its computations may often
lead to ties in the results, which is an intrinsic situation of exact
computations with the binomial distribution. However, if computations
are carried on with a clever implementation that avoids numerical
issues, then ties should arguably be considered as effective ties,
and other procedures/ideas could be devised to break the ties. For our
current needs, ties have not constituted a problem, mainly because we
are usually interested in the top confidence values, where the ties
seem to be less often.

\section{Schizophrenia data}
\label{sec4}

\begin{figure}[ht]
\centering
\subfloat[Transcript 204434.]{\includegraphics[width= 2.5in]{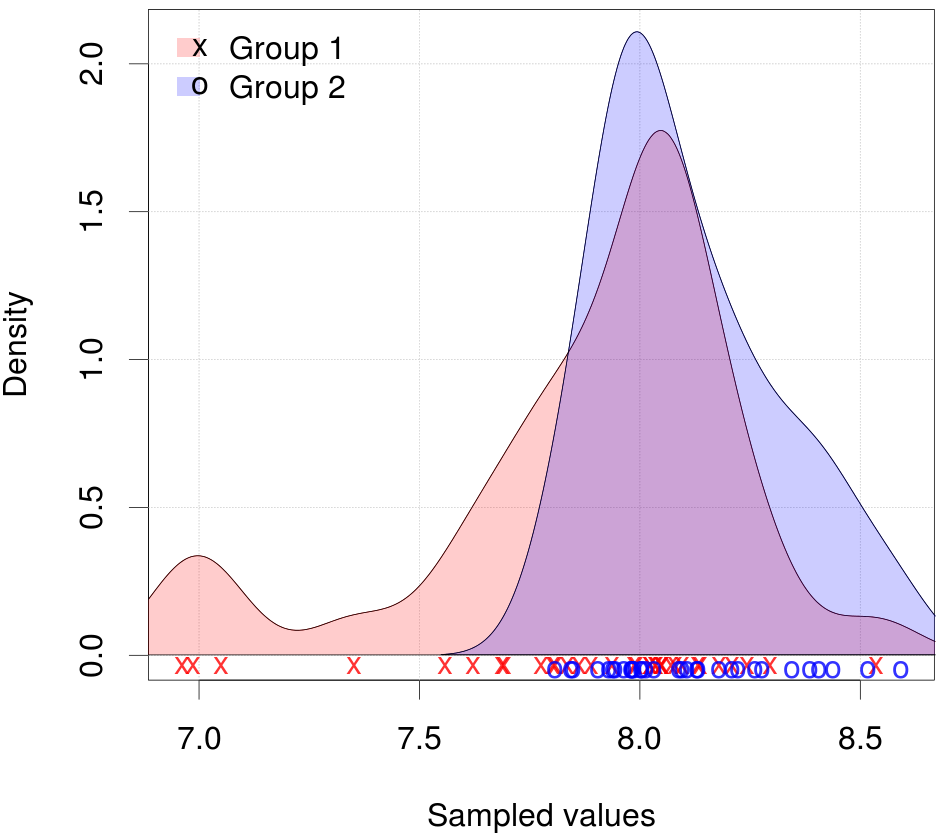}\label{f6a}}
\subfloat[Transcript 209847 (7th highest Quor Confidence).]{\includegraphics[width= 2.5in]{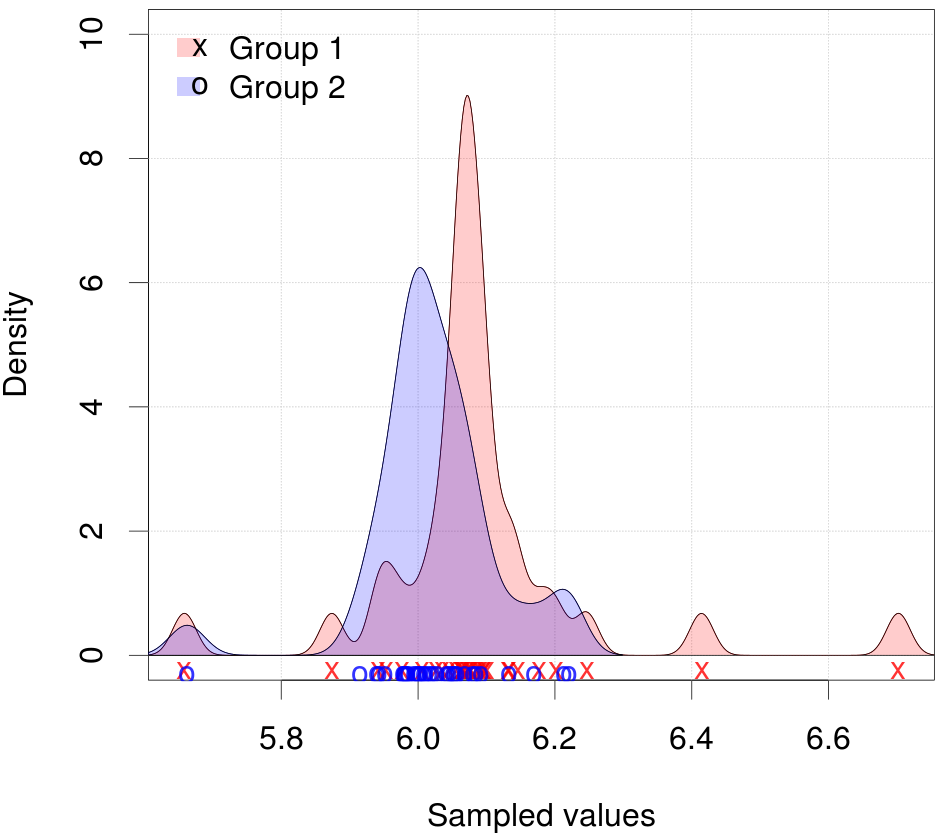}\label{f6b}}\\
\subfloat[Transcript 215001.]{\includegraphics[width= 2.5in]{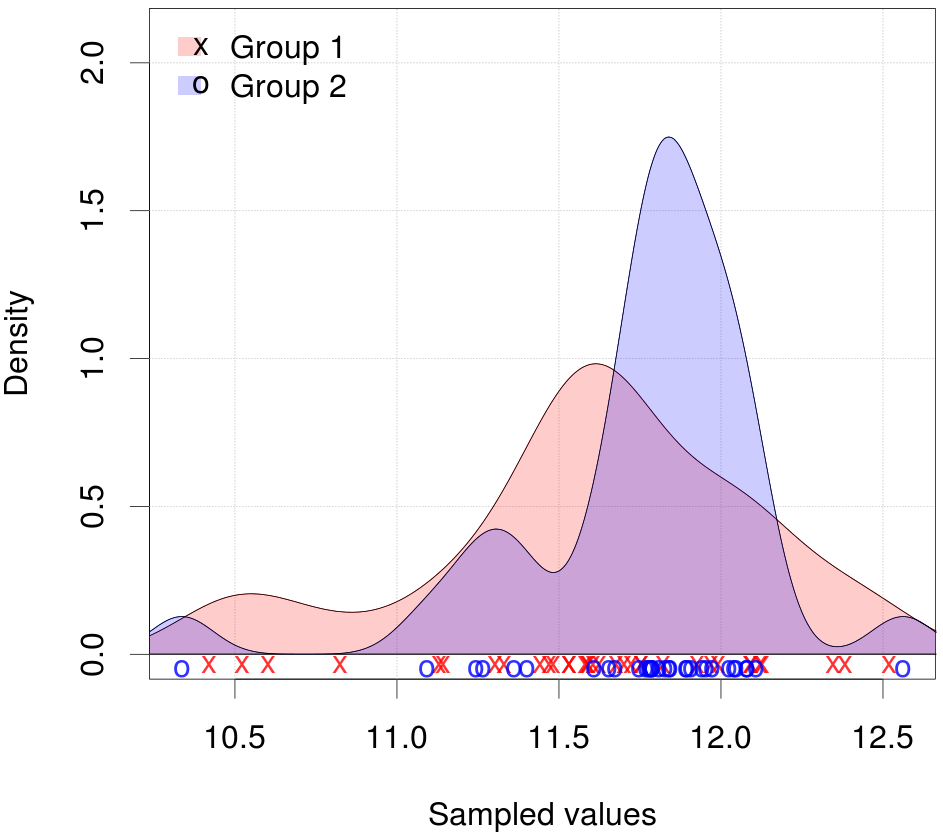}\label{f6c}}
\subfloat[Transcript 215003 (highest Quor Confidence).]{\includegraphics[width= 2.5in]{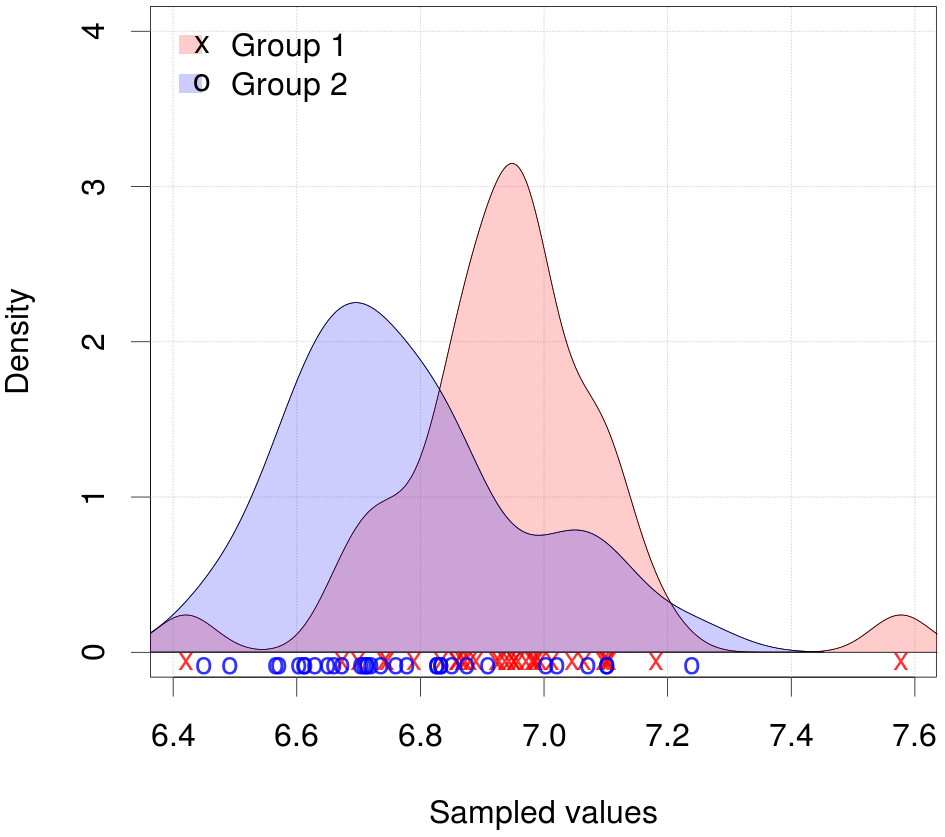}\label{f6d}}
\caption{Data of four transcripts from the Schizophrenia data set. In
  (a), Quor Conf=0.43, but $t$ and $u$ reject the null hypothesis at 0.002 and
  0.02 (that is, they consider the difference between groups to be quite significant). In (b), $t$
  cannot reject the null hypothesis, and in (c) neither $t$ nor $u$
can (in both cases Quor Conf is at least 0.95). In (d), both tests and Quor strongly agree
(Quor Conf=0.99, $t$-test pvalue=8e-4, $u$-test pvalue=2e-4).}\label{f6}
\end{figure}

We show a practical scenario by analyzing
the Schizophrenia data set from the Stanley Medical Research
Institute's online genomics database~\cite{higgs2006,stanley2012}. The sample sizes are $m_1 = 34$  individuals in the
control group and $m_2 = 34$ patients with schizophrenia. A total
of $20992$ microarray probes were obtained. The usual goal in such an
analysis is to find the most differentially expressed genes. For that
purpose, we decided to evaluate the confidence of the statements
$\{Q_1 < Q_2\}$ and $\{Q_1 > Q_2\}$, where $q_1=q_2=1/2$, that is, we
compared the medians of the populations. This is performed for each
gene in the data set. We argue that the difference in the populations'
medians indicate that genes might be differentially expressed (at
least it shows that the distributions are not equal). Not surprisingly,
there are {\bf no} significant genes if we perform either
$t$-test or $u$-test with multiple-test correction (we have used
the Holm-Bonferroni correction). On the other hand, Quor measure has a clear
interpretation as the confidence of the difference in medians, so if one chooses
genes with, for instance, confidence above 95\%, then there are 56 genes in this
situation, of which 14 cases suggest that ill patients (group
2) have under-expressed genes ($\{Q_1 > Q_2\}$) and 42 cases suggest
over-expressed genes ($\{Q_1 < Q_2\}$). 
\begin{table}[ht]
\begin{center}
\caption{Number of rejected and non-rejected genes among those with
  Quor confidence $\geq 95\%$ in the
  Schizophrenia data. First line is performed without
  using multiple test correction for $t$ and $u$ tests. Second does
  correction assuming that only 56 hypothesis tests are executed. If
  one corrects for all the genes, no significant gene is found by $t$
  or $u$ tests.}
\label{table3}
\begin{tabular}{c|cc|cc}
\hline
& \multicolumn{2}{c|}{$t$-test} & \multicolumn{2}{c}{$u$-test} \\
& Non-Reject & Reject &  Non-Reject & Reject\\
\hline
no correction & 18  &  38 & 1 &   55\\
with correction & 48 & 8 & 43 & 13\\
\hline
\end{tabular}
\end{center}
\end{table}

Because of its characteristic, one might see a single Quor confidence
as a more conservative approach than other tests, as it has fewer
assumptions (for instance, all but one high-confidence genes from Quor
also had $u$-test pvalue $<5\%$), but this vanishes if multiple test correction is
performed. Table~\ref{table3} shows the number of null hypothesis (that is, no difference) rejections
with $t$-test and $u$-test assuming that only 56 tests were performed
(exactly for those genes with high Quor confidence). This illustrates a
hypothetical procedure where Quor is firstly executed, followed by the
hypothesis testing (we decided to show this view otherwise those tests
would not be able to reject the null hypothesis for any gene). In the table we see
that $t$ and $u$ tests turn out to be very conservative because of multiple test
correction, and the prior use of Quor can alleviate this situation
and consequently reduce an excessive amount of Type II errors.

\section{Evaluation as Feature Selection}
\label{sec5}

 In this section we apply different hypothesis testing methods in the literature and
Quor to the task of feature selection, with the ultimate goal of
building a classifer with a subset of all variables. We do not compare these
approaches with yet other methods for feature selection, for example those which
consider correlations between variables and information measures, because they
have very different goals, while ours is to compare methods with similar
purpose, such as $t$-test, $u$-test, $ks$-test, and Quor itself. For that purpose,
we try to predict a yet unseen class/group given the observations of the model
covariates. In each data set, the class has a particular meaning, and
the number of samples and variables vary. Table~\ref{t0} shows the
main characteristics of the data sets with which we work. Please refer
to the corresponding citation for additional information on the
data~\cite{lymphomadata,colondata,leukdata,higgs2006,ovardata,prostatedata,breastdata,lungdata}. These
data were obtained from internet
repositories~\cite{repo2,seville,stanley2012}. 
\begin{table}[ht]
\begin{center}
\caption{Data set characteristics. $m$ is the number of patients,
  shown by the amount in each group. PS stands for {\it Proteomic
    spectra}, GEP for {\it Gene Expression Profiling}.}
\label{t0}
\begin{tabular}{c|cccc}
\hline
Data set & $m$ & \# Feat. & Type & Class \\
\hline
Prostate C.& 8+13 & 12600 & GEP & Relapse/Not \\
Schizophr. & 34+32 & 20992 & GEP & Disease/Not\\
Lung C.&  24+15 & 2880 & GEP & Relapse/Not \\
Breast C.             & 12+7 & 24481 & GEP & Relapse/Not\\
Colon C.   & 22+40 & 2000 & GEP & Disease/Not\\
Lymphoma    & 22+23 & 4026 & GEP & Dis. Subtype\\ 
Leukemia    & 27+11 & 7129 & GEP & Dis. Subtype\\ 
Ovarian C.    &  162+91 & 15154 & PS & Disease/Not\\
\hline
\end{tabular}
\end{center}
\end{table}

The idea is to fix the
classifier while varying the feature selection approach among Quor
(median as the chosen quantile), $t$-test, $u$-test and $ks$-test
(Kolmogorov-Smirnov test~\cite{kolm}). To avoid a result that is
specific to a single classifier, we perform the experiment with two
very distinct classifiers: a Bayesian network and a C4.5 decision tree
(both inferred from data), implemented in the Weka
package~\cite{weka}. For each data set, we run a five-fold
cross-validation procedure repeated 20 times (so as the test set is
never available during training; five folds are chosen because some of the
data sets contain very few patients). Results are shown in
Tables~\ref{tableBA} and~\ref{tableJ48} for the Bayesian network and
the decision tree, respectively. Numbers represent average accuracy
over $5\cdot 20=100$ runs. Quor, as a
feature selection procedure, has different characteristics from the
$t$-test, and each of them seems to perform better in distinct data
sets; on the other hand, when compared to the $u$-test, Quor seems to
show a mild improvement in accuracy.  This might be explained by the
greater proximity of ideas, in some sense, of the $u$-test and Quor,
with the latter demonstrating a slight better performance in these
specific data sets.

\begin{table}[ht]
\begin{center}
\caption{Average accuracy (over 100 runs) of a Bayesian network classifier using the 20
  best ranked features for each selection method.}
\label{tableBA}
\begin{tabular}{c|cccc}
\hline
Data set & Quor & $t$-test & $u$-test & $ks$-test \\
\hline
Prostate C.&  {\bf 47.85} &   43.55 &   42.45 &  42.60 \\
Schizophrenia & 59.08  &  57.99  &  58.98 &  {\bf 60.09} \\
Lung C.&  66.39 &  65.54  &   65.45 &  {\bf 66.52} \\
Breast C.             &  69.50 & {\bf 74.75} &  64.08 &   66.67  \\
Colon C.   & {82.87} &  76.60 &   {\bf 83.37}  & {81.84}  \\
Lymphoma    &  91.33 &   {\bf 92.44}  &   92.33   &  91.67  \\
Leukemia    &  {\bf 96.23} &   91.54 &  {96.07}   &  92.48 \\
Ovarian C.    &  97.41 & {\bf 98.00} &  97.23 &    97.35  \\
\hline
Average & {\bf 76.02} & 75.52 & 75.56 & 75.55 \\
\hline
\end{tabular}
\end{center}
\end{table}
\begin{table}[ht]
\begin{center}
\caption{Average accuracy (over 100 runs) of a C4.5 decision tree classifier using the 20
 best ranked features for each selection method.}
\label{tableJ48}
\begin{tabular}{c|cccc}
\hline
Data set & Quor & $t$-test & $u$-test & $ks$-test \\
\hline
Prostate C. & {40.45} & {\bf 40.70}        & 39.55         & 39.90          \\
Schizophrenia & 55.91 & 55.24          & 56.81    & {\bf 56.87}          \\
Lung C. & 60.02 & 59.07    & 63.09        & {\bf 64.27} \\
Breast C. & 50.67 & {\bf 59.50} & 51.67     & {58.33}\\
Colon C. & {\bf 83.65} & 77.40 & 80.57 & 81.01\\
Lymphoma & 76.89 & {\bf 77.44}      & 77.11        & {\bf 77.44}       \\
Leukemia & {\bf 87.66} & 84.70     & 86.55        & 86.11       \\
Ovarian C. & 96.68 & {\bf 97.21} & 96.66     & 96.58   \\
\hline
Average & 68.27 & 68.23  & 68.74      & {\bf 69.76}   \\
\hline 
\end{tabular}
\end{center}
\end{table}

\section{Conclusions}
\label{sec6}

This study proposes a method called Quor to compute the confidence of
a statement about the order of arbitrary quantiles of many
populations.  Its only assumption is independence of samples, which
makes it applicable to many domains. Among Quor properties, we
highlight its nonparametric nature, the possibility of processing data
sets with missing data and varying number of samples from each
population (across multiple tests), the confidence interpretation that
might avoid the need of multiple-test correction, and its efficient exact
computations.  Quor will be made available under an open source
license both as an easy-to-install R package and as a plugin for the
Weka Data Mining package~\cite{weka}. Besides confidences on the order
of quantiles of the populations, the package also computes confidence
intervals based on order statistics (we refrained from detailing on
them here; see Chapter 7 of~\cite{david2003}).

Our empirical analysis has indicated that Quor results are in line
with the most used techniques for hypothesis testing, which are also often used
to order genes according to their statistics/p-values. Quor also performs
well as a feature selection procedure (when compared to hypothesis
testing methods) in a benchmark of high-dimensional data sets.  As
future work, we will study the occurrence of ties in the results of
Quor, and means to break such ties. While they have not been a major
issue so far in our analyses, we consider important to have a sound
procedure to resolve them, especially to deal with discrete ordinal
data. We are also pursuing applications of the method in recent data
sets from the latest microarray technologies. As the number of
covariates increases as Quor shall become important for its
computational performance and minimal assumptions.

\section*{Acknowledgements}

This work was partially supported
by the Brazilian agencies FAPESP, CNPq and CAPES.

\bibliographystyle{plain}
\bibliography{document}
\end{document}